\documentclass[twocolumn,showpacs,amsmath,amssymb]{revtex4}
\usepackage{graphicx}
\usepackage{dcolumn}
\usepackage{bm}
\usepackage[dvips]{epsfig}
\newcommand{\ba}{\begin{eqnarray}}
\newcommand{\ea}{\end{eqnarray}}
\newcommand{\ban}{\begin{eqnarray*}}
\newcommand{\ean}{\end{eqnarray*}}
\newcommand{\be}{\begin{equation}}
\newcommand{\ee}{\end{equation}}
\newcommand{\bd}{\begin{displaymath}}
\newcommand{\ed}{\end{displaymath}}

\newcommand{\n}[1]{\label{#1}}

\newcommand{\eq}[1]{(\ref{#1})}

\newcommand{\hF}{\hat{\Phi}}

\newcommand{\hh}{\, ,\hspace{0.5cm}}
\newcommand{\hhh}{\, ,\hspace{0.2cm}}

\newcommand{\APJ}{ApJ}
\newcommand{\APJl}{ApJ}

\newcommand{\mnras}{MNRAS}
\newcommand{\na}{New A}

\newcommand{\pasj}{PASJ}

\begin{document}

\draft


\title{Ray-tracing in four and higher
dimensional black hole spacetimes:\\
An analytical approximation}
\author{Patrick Connell}
\email{pconnell@phys.ualberta.ca}
\author{Valeri P. Frolov}
\email{frolov@phys.ualberta.ca}
\affiliation{Theoretical Physics Institute, University of Alberta,
Edmonton, Alberta, Canada, T6G 2J1}

\date{\today}

\begin{abstract}
We study null rays propagation in a spacetime of  static Schwarzschild--Tangherlini
black holes in arbitrary number of dimensions. We focus on the bending
angle and the retarded time delay for rays emitted in the vicinity of
a black hole and propagating to the infinity. We obtain an analytic
expression in terms of elementary functions which approximate the
bending angle and time delay in these spacetimes with high
accuracy. We analyze the relative error of the developed analytic
approximations and show that it is quite small in the complete domain
of the parameter space for the rays reaching the infinity and for
different number of the spacetime dimensions. Possible applications of
the obtained results are briefly discussed.
\end{abstract}

\pacs{04.70.Bw, 04.50.-h, 04.25.-g, 04.20.Jb \hfill  
Alberta-Thy-04-08}

\maketitle

\section{Introduction}

Effects predicted by the general relativity are important in the vicinity
of compact objects, such as neutron stars and black holes. One can
study such effects by observing electromagnetic or gravitational
radiation from these objects. The first light detected from regions
close to the black holes was discovered by the ASCA satellite,
\cite{Ta,Ya}. Now there is much evidence that interesting
astrophysical effects are connected with the physics in the vicinity
of black holes. Broadening of X-ray emission lines from accretion
disks, 
\cite{Reynolds,Turner,Nandra,Fanton,Bromley,Dabrowski,Cadez,Martocchia},
and X-ray flares, \cite{Baganoff,Goldwurm}, or quasi-periodic
oscillations, \cite{Strohmayer,Miller}, are examples of these
effects. To explain these observations one needs not only to develop
models of these phenomena but also to solve equations for the light
propagation in the gravitational field of these objects.

If the wave-length of the radiation is much smaller than
the characteristic scale ($r_{g}$, the gravitational radius) one can use the
geometric optics approximation and reduce the problem to the study
of null ray propagation in curved spacetime. Two related
problems are of interest: (1) how does a distant observer see objects
located far `behind' a black hole distorting images; (2) how is
the light emitted by matter in the region of a strong gravitational
field seen by a distant observer. For example, the radiating matter
can be the surface of a collapsing body, or the surface of a neutron star,
or an accretion disk and so on. A similar problem is of potential
interest for higher dimensional black holes. In brane models with
large extra dimensions mini black holes can play a role of `probes'
of extra dimensions. Scattering or emission of light or gravitons
by such black holes might be interesting in this connection (a recent
review on higher dimensional black holes can be found in \cite{EmRe},
see also references therein). 

There exist a lot of publications where ray-tracing in Schwarzschild
and Kerr geometries has been discussed in detail. It
is well known that the Hamilton-Jacobi equations for light rays in
these geometries allow separation of variables. This property is
connected with the existence of an additional quadratic in momentum
integral of motion associated with the Killing tensor,
\cite{Carter:68}. In the Schwarzschild 4D geometry expressions for
scattering data (such as a bending angle and time delay) can be
written explicitly in terms of elliptic integrals, \cite{Darwin:59},
\cite{Darwin:61}. In the Kerr metric such quantities can be expressed
in terms of generalized hyper-geometric functions, \cite{Kran}.

Recently it was shown that higher dimensional rotating black holes in
many aspects are similar to the 4D Kerr black holes. Namely, the most
general Kerr-NUT-(A)dS metric always possesses a so-called {\em
principal conformal Killing-Yano tensor} \cite{FrKu,KuFr} which generates a set
of the second rank Killing tensors \cite{Kill_1,Kill_2}. As a result
the geodesic equations in such spaces are completely integrable
\cite{Kill_2,Kill_3}, and the Hamilton-Jacobi, Klein-Gordon and Dirac
equations allow complete separation of variables \cite{FrKuKr,Oot}. It
means, that a solution of the geodesic equations can be obtained in
quadratures. However, even for the non-rotating black holes in
higher dimensions a solution cannot be expressed in terms of known
special functions (for a general discussion of hidden symmetries and separation
of variables in higher dimensional black holes see recent reviews
\cite{Fr,FK08} and references therein).  

For scattering at small angles the integrals arising in these
problems can be estimated by using the perturbation theory (see e.g.
\cite{Hobill}). One can use also numerical calculations, as it was
done in \cite{GoFr} where the capture cross-sections for a
five-dimensional rotating black hole was studied. Usually the
ray-tracing, used for the reconstruction of spectrum and light curves
as seen at infinity for light emitting patterns, requires
repeating calculations for very many rays. To reduce the time and
cost of the calculations, and to be able to study analytically
different qualitative observed features of the radiation it is useful
to have simple analytical expressions, for example in terms of
elementary functions, which approximate accurately enough the
scattering data. In this paper we develop a scheme for obtaining such
approximations for four and higher dimensional non-rotating black
holes. 

An elegant analytical formula approximating bending angles for null
rays in the 4D Schwarzschild geometry was proposed by Beloborodov,
\cite{Belo:02}. The accuracy of the Beloborodov's approximation for
light rays passing the gravitating object at $R=3r_g$ is of the order
of 1\%, but unfortunately for the rays passing closer to the black hole its
accuracy becomes worse and reached, e.g., 10\% for those rays which
pass at $R=2r_g$. In the more recent paper \cite{FroLe:05} there was
proposed a better analytical approximation for the bending angle and
time delay . It has an accuracy of $2-3$\% for rays emitted at the
radius $R\ge 2r_g$. However, for rays emitted inside $R=4M$ the
accuracy of this approximation diminishes. The reason is this: the
bending angle for rays with impact parameter close to the critical
value $\lambda_{*}=3\sqrt{3}M$ passing close to the critical radius
$r_{*}=3M$  becomes (logarithmically) large.  The asymptotic behavior
of the bending angle for near critical rays was studied in
\cite{Esh}.  Very recently  an analytical
procedure for studying gravitational lensing in the strong deflection
regime  was proposed, \cite{Bozz}. In essence \cite{Bozz} extracts from the bending the angle
the divergent term and studies the behaviour of the deflection of
light near the critical radius. As a result \cite{Bozz} constructs an
approximation to the four dimensional gravitational lensing equation
in the strong deflection limit for the Kerr black hole and spacetimes
which have a stationary spherically symmetric line element. In the Schwarzschild case
for photons emitted at $R=4M$ and scattered to infinity at $R=3.05M$, for example, the method
outlined in \cite{Bozz} results in a relative error for the bending angle on the order of $0.5\%$
when the impact parameter has the value $5.198M$, and for those photons escaping to infinity
from inside the critical radius at $R=2.5M$, for example, the relative error is of order $0.05\%$ when the impact parameter is  
$5.144M$.

In the present work we propose an improved analytical approximation
for the ray tracing problem in four and higher dimensional static
black hole space-times. We analyze the following problem: suppose a
ray is emitted at the radius $r_0$ with impact parameter $\lambda$.
We obtain an analytic expression for the bending angle and time delay
for such rays which is uniformly valid for the two parameter set
$\{r_0,\lambda\}$ specifying the ray. To make an approximation
possible we first extract from the integrals for scattering data the
contributions which are logarithmically divergent near the critical
trajectories.  The key observation of the present work is that after
this procedure the remaining part can be approximated with very high
accuracy by a function of one variable. By finding a proper approximation
for this function one can obtain a very accurate uniform
approximation for the required quantities.

The paper is organized as follows. In section~II we collect formulas
for the ray propagation in the Schwarzschild and Tangherlini
geometries. In sections~III and IV we derive the expressions which we
use to approximate  the bending angle and time delay, and study the
errors of the approximation. Section V contains a discussion of the
obtained results and their possible applications.

\section{Null rays in a static black hole geometry}

\subsection{Basic equations}

We consider null rays propagating in the background  of a
$D$-dimensional Schwarzschild-Tangherlini metric
\be\n{metric}
ds^2 = -fdt^2 +f^{-1}dr^2 +r^2d\Omega^2_{n+1}\, ,
\ee
where $n=D-3$,
\be
f=1-\left(r_{g}\over r\right)^n\, ,
\quad
r^n_{g}\equiv \frac{16{\pi}M}{(n+1)A_{n+1}}\, .
\ee
Here $M$ is the mass of the black hole and $d\Omega^2_{n+1}$ is a line
element on a unit $(n+1)$-dimensional sphere, $S^{n+1}$,
\be
d\Omega^2_{1}=d\theta_0^2\equiv d\phi^2\hh
d\Omega^2_{n+1}=d\theta_n^2+\sin^2\theta_n d\Omega^2_{n}\, ,
\ee
\[
\theta_{i\ge 1}\in [0,\pi]\, ,\phi\equiv\theta_0\in [0,2\pi]\, ,
\]
and $A_{n+1}$ is the area of $S^{n+1}$,
\ba
A_{n+1}=\frac{2\pi^{(n+2)/2}}{\Gamma(\frac{n+2}{2})}\, .
\ea
Here $\Gamma (z)$ is the Euler Gamma function. We use units in which the $D$-dimensional gravitational coupling
constant $G$ and the speed of light $c$ are equal to 1.

It is easy to show that similar to the $D=4$ case a photon trajectory
in the spacetime \eq{metric} lies always within a plane.  Without loss of
generality we assume this plane to be the equatorial plane, i.e.
$\theta_{i}=\pi/2$, $i=1,..,n$. We choose an affine parameter
$\zeta$ so that photon's  $D$-momentum is $p^{\mu}=dx^{\mu}/d\zeta$
and for our choice of the coordinates we have
\ba
p^{\mu}=(p^t,p^r,p^{\phi},0,\ldots,0)\, .
\ea
The energy $E=-p_{t}$ and the angular momentum
$L=p_{\phi}$ are integrals of the motion.
Since a trajectory with $L<0$ can be obtained from a trajectory $L>0$ by
a simple reflection $\phi\to -\phi$ we assume that $L\ge 0$.
Using the integrals of motion and the relation $p_{\mu}p^{\mu}=0$ one
can write the equations of motion in the form
\ba\n{r}
\dot{r}&=&\sigma_r E\, Z\, ,\quad
Z=\sqrt{1-{\lambda^2}f/r^2}\, ,
\\
\n{t}
\dot{t}&=&E/f\, ,\quad
\dot{\phi}=L/r^2\, .
\ea
Here the dot over an expression means its derivative with respect to
the affine parameter $\zeta$ and  $\sigma=\pm 1$.  For the
outward moving photon $\sigma_r=1$, while for the inward moving one
$\sigma_r=-1$. A change of the sign of $\sigma_r$ occurs at a turning
point $r_*$ defined by the relation 
\be
1-{\lambda^2}f_*/r_*^2=0\,,
\quad f_*=f(r=r_*)\, .
\ee
By excluding the affine parameter the equations \eq{r}-\eq{t} can be
written in the form
\ba
\frac{d\phi}{dr}=\frac{\sigma_r\lambda}{r^2Z}\hh
\frac{dt}{dr}=\frac{\sigma_r}{f\, Z}\, ,
\ea
where  $\lambda=L/E$ is the photon's  impact parameter.

Consider a photon emitted at point $r=r_0$. We call such a photon
forward (backward) emitted if $\sigma_{r_0}>0$ ($\sigma_{r_0}<0$) at the point
of emission. For a photon propagating from $r_0$ to infinity we define a
bending angle as $\Delta\phi=\phi(\infty)-\phi(r_0)$. The bending
angles for the forward and backward emitted photons are, respectively,
\ba
\Delta\phi_{+}=\Phi(r_0)\, ,\quad
\Phi(r_0)=\int^{\infty}_{r_0} \frac{\lambda dr}{r^2
Z}\, ,\\
\Delta\phi_{-}=\left[\int^{r_0}_{r_*}
+\int^{\infty}_{r_*}\right] \frac{\lambda dr}{r^2 Z}=2\Phi(r_*)-\Phi(r_0)\, .
\ea

\begin{figure}[tp]
\begin{center}
\includegraphics[height=1.2in]{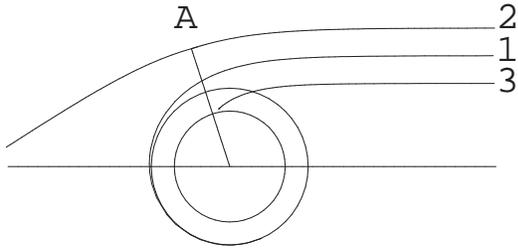}
\caption{The above figure shows qualitatively the different types of
motion for light rays in a spherically symmetric black hole
background. Line 2 shows a ray which has a turning point, A.
It corresponds to the scattering problem, that is when the ray 
coming from infinity is directly
scattered back to infinity. Ray 1 corresponds to a critical trajectory
when an incoming photon  spirals 
around the circle (the critical radius) an infinite number of
times. Ray 3 describes a motion of a photon coming from infinity which
is captured by the black hole.}
\label{fig1}
\end{center}
\end{figure}

Similarly, the time of arrival of a forward emitted
photons as measured by an observer at radius $r_1$ is
\ba\n{to}
t^{(o)}_{+}=t^{(e)}_{+} + \int^{r_1}_{r_0}
\frac{dr}{f\,Z}\, .
\ea
The integral in \eq{to} is divergent when the upper limit tends to
infinity. This
divergence reflects a simple fact: namely, the time required for a light ray
to reach infinity is infinitely large. For this reason it is more
convenient to deal with the retarded time
\be\n{tort}
u=t-\hat{r}\, \quad
d\hat{r}={dr\over f}\, . 
\ee
After simple transformations one has
\ba\n{uo}
u^{(o)}_{+}=u^{(e)}_{+} + T(r_0)\, ,
\ea
\be\n{TT}
T(r_0)=\int^{\infty}_{r_0}
\,\frac{dr}{f}\left({1\over Z}-1\right)\, ,
\ee
\be
u^{(o)}_{-}=u^{(e)}_{-} + 2(\hat{r}_{0}-\hat{r}_{*})+
2T(r_{*})-T(r_0)\, ,
\ee
where $\hat{r}_0$ and $\hat{r}_*$ are tortoise--like coordinates, \eq{tort},
for the point of emission and turning point, respectively.

In what follows we focus on the functions $\Phi$ and $T$ since the
bending angle and retarded time for both forward and backward emitted
photons can be expressed in terms of these quantities.  

In the 4
dimensional spacetime, $n=1$, the integrals $(10)$ and $(15)$ can be
written in terms of elliptic functions. In the higher dimensional
case the integrals cannot be written in terms of known special
functions. Our aim is to study these objects in a spacetime with
arbitrary number of dimensions as functions of two variables, the
impact parameter $\lambda$ specifying the photon trajectory, and the
point of emission, $r_0$. It should be emphasized that knowledge of
these scattering data allows one to obtain expressions for other
quantities which might be of interest. For example, if one wants to know
what are the bending angle and retarded time delay for a photon with given
impact parameter propagating from one point, $r_0$, to another,
$r_1$, it is sufficient to calculate the differences between the
corresponding quantities
$\Delta\phi_{\lambda}(r_1)-\Delta\phi_{\lambda}(r_0)$ and $\Delta
u_{\lambda}(r_1)-\Delta u_{\lambda}(r_0)$.

\subsection{The equations of motion in a dimensionless form}

The problem contains only one dimensional parameter $r_g$, which
determines the scale. It is convenient to introduce dimensionless
quantities
\ba
q\equiv\frac{r_{g}}{r}\, ,\quad   q_0\equiv\frac{r_g}{r_0} \, ,\quad
l=\frac{\lambda}{r_{g}}\, .
\ea
In these variables
\be
f=1-q^n\, ,\quad Z=\sqrt{1-l^2q^2\, f}\, .
\ee
The motion of the photon is possible only in the region where
\be
0\le l\le l_{max}={1\over q\sqrt{1-q^n}}\, .
\ee
(Notice that we consider only non-negative impact parameters.)
On a plot in $(q,l)$ variables this region lies below the line
$l_{max}(q)$. The allowed regions for different number of dimensions
($n\ge 1$) are similar. Figure~2 shows the allowed domain for $n=1$.  
The minimum of the function
$l_{max}(q)$ is at $q=q_*$, where
\be
q_*=\left( {2\over n+2}\right)^{1/n} .\ 
\ee
At this point $l=l_*$ and $f=f_*$ where
\be
l_*=\sqrt{ {n+2\over n}} \left( {n+2\over 2}\right)^{1\over n}\hh
f_*={n\over n+2}\, .
\ee

\begin{figure}[tp]
\begin{center}
\includegraphics[height=1.8in]{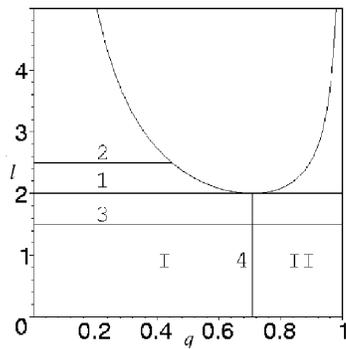}
\caption{Function $l_{max}(q)$.  Motion of photons is allowed in the
region below this curve. Points $q=0$ and $q=1$ correspond to the infinity
and the horizon, respectively. Horizontal lines describe the photon
trajectories. Line 2 corresponds to a photon scattering, while line 3
corresponds to the photon capture. Line 1 is critical. It corresponds
to a photon with the critical impact parameter $l_*$. Domain $I$ is a
region below $=l_{max}(q)$ lying to the left from a vertical line 4. A
region to the right from line 4 and below line 1 is domain $II$. In what
follows we consider only photons emitted from domains $I$ and $II$, which
can reach the infinity.}
\label{fig2}
\end{center}
\end{figure}

Before going further we distinguish the possible qualitatively
different types of the light ray trajectories (see Fig.~\ref{fig2}). 
Rays with $l<l_*$ can  propagate from infinity ($q=0$) to the horizon
($q=1$) or travel to infinity from a point $q$ satisfying $q < 1$. Rays with $l=l_*$ can either approach $q=q_*$ coming from
the black hole ($q>q_*$)  or from infinity $q=0$). In doing so they
spiral around $q=q_*$ an infinite number of times. Rays with $l>l_*$
have a turning point. They can either propagate from infinity to some
periastron and return to infinity or from the black hole horizon to
an apastron and after fall back to the black hole. The latter case is
not of much interest for astrophysical and other applications and we
shall not consider it. 

Using the above introduced dimensionless quantities we can rewrite
the expressions for the bending angle and time delay in the following
form
\ba\n{bend}
&&\Phi=\int_0^{q_0}{l\, dq\over Z(q)}\, ,\\
&&T=r_g\int_{0}^{q_0}\frac{dq}{q^2f}\left({1\over
Z(q)}-1\right)\n{time}\, ,\\
&&Z^2(q)=1-l^2q^2(1-q^n)\, .
\ea

Let us consider the function $Z(q)$ in the interval from $q=0$
(infinity) to $q=1$ (horizon).
It monotonically decreases from its value 1 at $q=0$
until the minimum
\be
Z_*=\sqrt{1-(l/l_*)^2}\,
\ee
at $q=q_*$, and then monotonically increases until it has the value 1 at
$q=1$.
It means that one can expect that for the integrals in \eq{bend} and
\eq{time} the main contribution comes either from the region near the point
$q_*$ if it enters the integration domain, $q_*\in (0,q_0)$, or
from the region near the end point $q_0$ in the opposite case. We use
this remark to construct an approximation for these integrals.

\section{Approximating the bending angle}

\subsection{Type $I$ rays}
\n{B1}

\subsubsection{Leading part}

First we study the rays emitted from the domain $I$ ({\em type $I$
rays}). Let us introduce a new coordinate $y$
related to $r$ and $q$ as follows
\be
y=1-r_0/r\hh q=q_0(1-y)\, ,
\ee
and denote
\ba\n{pp}
p_{0}&\equiv& q_{0}/q_{*}\hhh
p\equiv q/q_{*}=p_0(1-y)\hhh
\, ,\\
P&=&p_0^n\hhh B=\nu^2\hh\nu\equiv l/l_{max}\, .
\ea
By fixing the parameters $P$ and $B$ one uniquely specifies a ray and
its emission point. For the rays of type $I$ one has $0\leq P\leq 1$
and $0\leq B\leq 1$.

Using these notations one gets
\be
f=1-{2P\over n+2}(1-y)^n\, ,
\ee
and the expression for the bending angle for
the forward emitted photons takes the form
\ba\n{Phi}
\Phi &=&\int^{1}_{0}\frac{\nu\, dy}{Q}\, ,\\
Q &=&\sqrt{f_0}Z=\sqrt{f_0-\nu^2(1-y)^2\, f}\, .
\ea
One also has
\ba
Q^2|_{y=0}&=&(1-B)(1-{2P\over n+2})\, ,\\
Q^2|_{y=1}&=&f_{0}=1-2P/(n+2)\, .
\ea

We study now $\Phi$ as a function of $(P,B)$
in the domain $I:$\ \ $0\le  P, B\le 1$. Our goal is to obtain
an analytic expression uniformly approximating $\Phi$ in this region.

For the interval $(0,q_0\le q_*)$ in the domain $I$ the function $Q$
has its minimal values at $y=0$ which corresponds to $q=q_0$.  We
denote by $\hat{Q}^2$ the  expansion of $Q^2$ in $y$ up to the second
order
\ba\n{QQ}
\hat{Q}^2&=&a+2by+cy^2\, ,\\
a&=&(1-B)(1-{2P\over n+2} )\, ,\nonumber\\ b&=&B(1-P)\, ,\\
c&=&-B[1-(n+1)P]\, .\nonumber
\ea
In the domain $I$ the parameters $a$ and $b$ are always
positive, while $c>0$ for $P>1/(n+1)$ and $c<0$ for $P<1/(n+1)$. At the
point $y=0$, $\hat{Q}^2$ and its  first two derivatives coincide with
the similar quantities for the  function $Q^2$. In particular, at in
the vicinity of the point $(P=B=1)$ one has
\be\n{crit}
Q\sim \hat{Q}\sim \sqrt{n}y+\ldots \, .
\ee
At the point $y=1$ one has
\be\n{AAA}
A\equiv \hat{Q}^2|_{y=1}=1-{2P\over n+2}+{n(n+1)\over n+2}BP\, .
\ee

Let us denote
\ba\n{I}
\hat{\Phi}=\int^{1}_{0}\frac{\nu\, dy}{\hat{Q}}\, ,
\ea
and present $\Phi$ in the form
\be\n{Delta}
\Phi=\hat{\Phi}+\Delta{\Phi}\, .
\ee
Both integrals, $\Phi$ and $\hat{\Phi}$ are logarithmically divergent
at the lower limit, $y=0$, for the point $P=B=1$ of the parameter
space. This point corresponds to a trajectory with the impact
parameter $l=l_*$ emitted at $q=q_*$.  As a consequence of the property
\eq{crit} the divergences of the both quantities $\Phi$ and $\hat{\Phi}$ are
identical and hence $\Delta \Phi$ remains finite at $P=B=1$.
The integral \eq{I} can be easily taken. We shall discuss now its
exact value and later we shall find a suitable approximation for
$\Delta{\Phi}$ valid in the total domain $I$.

\begin{figure}[tp]
\begin{center}
\includegraphics[height=2in,width=2.5in]{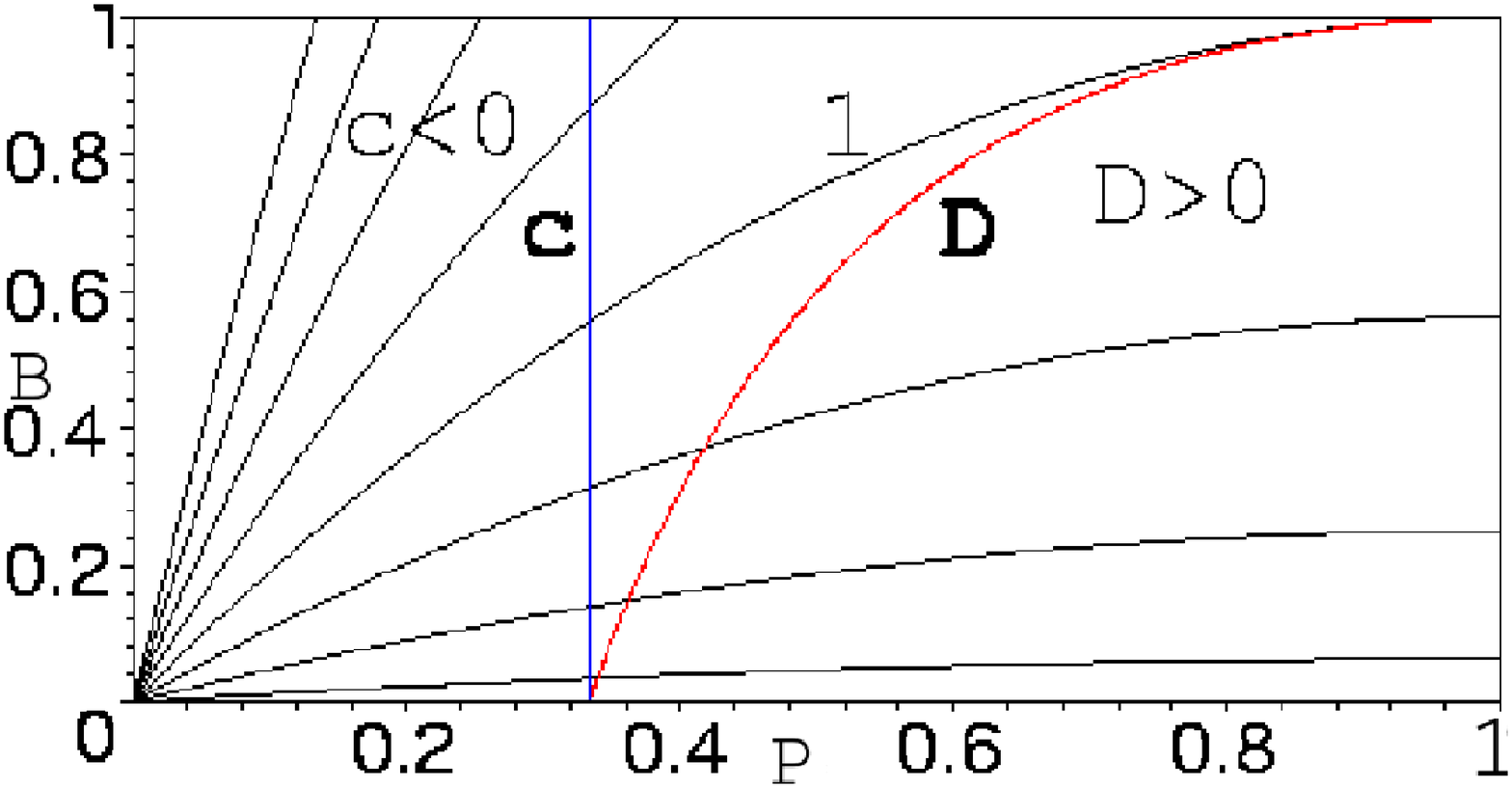}
\caption{The above figure shows the domain $I$ of  Figure~2 in
$(P,B)$ coordinates.  The line 1 originating from the point $(0,0)$
and terminating at the point $(1,1)$ corresponds to photons with
the critical value of the impact parameter. Lines originating from
the point $(0,0)$ which are above line~1 correspond to line~2 in
figure~2 and lines below it correspond to line~3 in figure~2. The
line $\bf c$ where $P=1/(n+1)$,  separates two regions where $c<0$
(to the left) and $c>0$ (to the right).  The
discriminant $D$ vanishes on the line ${\bf D}$.  A region with
negative $D$ is to the left of this line, and the region with positive
$D$ is to the right of the line ${\bf D}$. The plot is shown for $n=2$.
For other values of $n$ the structure is similar.}
\vspace{0.4cm}
\label{fig3}
\end{center}
\end{figure}

The form of the expression for $\hF$ depends on the sign of c and on
the sign of the discriminant $D$ (see figure~\ref{fig3}),
\be\n{DD}
D=ac-b^2\, .
\ee
The equation of the line where $D=0$ is
\be
B=[(n+1)P-1](n+2-2P)/(nP(n+1-P))\, .
\ee
This line is shown in Fig.~\ref{fig3}. It starts at $B=0$ and
$P=1/(n+1)$ and ends at the corner $P=B=1$. The vertical lines in the
figure~\ref{fig3} correspond to the fixed value of $P$ (and hence of
$q$ and $r$). The lines which start at the point $(0,0)$ in this
figure correspond to the fixed values of the impact parameter $l$ and
are described by the equation
\be
B=l^2 q_*^2 P^{2/n}[1-2P/(n+2)]\, .
\ee
Inside the domain $I$ for $c<0$ one has $D<0$ and
\ba\n{a}
\hat{\Phi}=
\frac{\nu}{\sqrt{{-c}}} \arcsin\left( {\sqrt{{-c}}B\over {D}}
[nP\sqrt{a}-(1-P)\sqrt{A}]\right)\, .
\ea
The parameters $A$ and $D$ which enter this relation are defined by
\eq{AAA} and \eq{DD}, respectively.

Taking the limit $c\to 0$, {\bf $(B\neq 0)$}, in this
expression one obtains 
\ba\n{b}
\hat{\Phi}=\frac{\nu}{b} (\sqrt{a+2b}-\sqrt{a})\, .
\ea
Finally for $c>0$ one has 
\ba\n{c}
\hat{\Phi}=\frac{\nu}{\sqrt{c}}\ln{\left(
\frac{nBP+\sqrt{cA}}{b+\sqrt{ac}}\right)}\, .
\ea

We shall need later the expression of $\hat{\Phi}$ for the
special  value $P=1$. In this case $c>0$. We denote this quantity by
$\hat{\Phi}_*$.  For this choice one has $q_0=q_*$, $l_{max}=l_*$
and  $\nu$ takes the value $\mu=l/l_*$. Taking this limit in the
formula \eq{c} one obtains
\be\n{PPP}
\hat{\Phi}_*={1\over \sqrt{n}} \ln \left[
{\mu\sqrt{n+2}+\sqrt{(n+2)\mu^2+1-\mu^2}\over
\sqrt{1-\mu^2} } \right] \, .
\ee
For a trajectory close to the critical one, $\mu\approx 1$, 
\be
\hat{\Phi}_*\sim -{1\over 2\sqrt{n}}\ln{(1-\mu)}\, .
\ee

\subsubsection{Approximating $\Delta\Phi$}

\begin{figure}[tp]
\begin{center}
\includegraphics[height=2.8in]{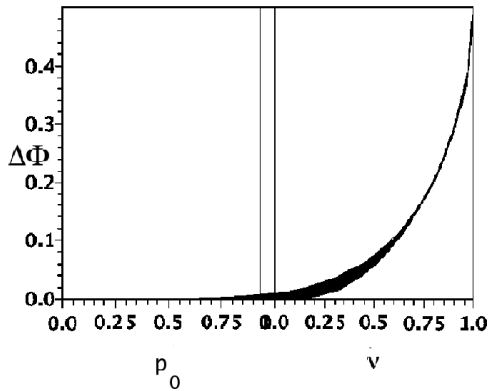}
\caption{This is a plot for $n=2$ of the surface $\Delta{\Phi}$ for
rays in domain $I$ when it is rotated to have an orientation
[-47$^{\circ}$, 90$^{\circ}$]} \vspace{0.4cm}
\label{fig4}
\end{center}
\end{figure}

Now we focus our attention on the quantity $\Delta{\Phi}$ which
describes the difference between the exact value of the bending angle
$\Phi$ and its leading part $\hat{\Phi}$ which we calculated in the
previous subsection. We have plotted $\Delta{\Phi}$ as a function of
$p_0$ and $\nu$ in the domain $I$. After a study of these plots we
found a remarkable fact. Namely, $\Delta{\Phi}$, which by its
definition is a function of 2 variables, $P$ and $B$, can be
approximated  accurately enough by a function of one variable, which
is a linear combination of $p_0$ and $\nu$. We found that this fact
is valid not only in 4 dimensions, but in higher dimensions as well.
Figure~\ref{fig4} illustrates this. It gives an example of the 3D
plot of $\Delta \Phi$ constructed by ${\it Maple}$ for $n=2$ ($D=5$) with the
orientation option for the view angles  chosen to be $[-47^{\circ},
90^{\circ}]$. The 2D surface in this projection is a slightly
broadened curve. Similar orientation angles can be found for other
dimensions. Using this fact  and by fitting the corresponding 2D
plots, one can find the following expression which approximate 
$\Delta \Phi$ with a very high accuracy
\ba\n{DP1}
\Delta \hat{\Phi}&=&-k_{\Phi}(\sqrt{1-(x_{\Phi}-1)^2}-1)
\Theta(x_{\Phi}-1)\, ,\\
x_{\Phi}&=&\sqrt{2}[\cos(\frac{\pi}{4}+\psi_{\Phi})
p_{0}+\sin(\frac{\pi}{4}+\psi_{\Phi})\nu] \n{DP2}\, .
\ea
Here $\Theta$ is the Heaviside step function.
The parameters $k_{\Phi}$ and ${\psi}_{\Phi}$ for different number of
spacetime dimensions are given in Table~\ref{tt1}.

\begin{table}[htb]
\begin{center}
\begin{tabular}{|l|c|c|c|c|c|}
\hline
Dimension & $k_{\Phi}$  & $\psi_{\Phi}$ & Percentage error
$\delta_{\Phi}$
\\[1.0ex]
\hline
D=4,n=1  & $0.47$ & $\pi/45$ &[-0.77\% , 1.78\%] \\[1.0ex]
\hline
D=5,n=2 & $0.49$  & $\pi/90$ &[-1.20\% , 1.69\%]  \\[1.0ex]
\hline
D=6,n=3 & $0.52$  & $0$ &[-2.10\% , 1.71\%]  \\[1.0ex]
\hline
D=7,n=4  & $0.55$ & $0$ &[-2.89\% , 2.58\%]  \\[1.0ex]
\hline
\end{tabular}
\end{center}
\caption{Approximation parameters $k_{\Phi}$ and
$\psi_{\Phi}$, and range of relative errors $\delta_{\Phi}$ for $\Phi$
in the domain $I$ for the spacetime of dimension $D=n+3$.}
\n{tt1}
\end{table}

\begin{figure}[tp]
\begin{center}
\includegraphics[height=2.5in]{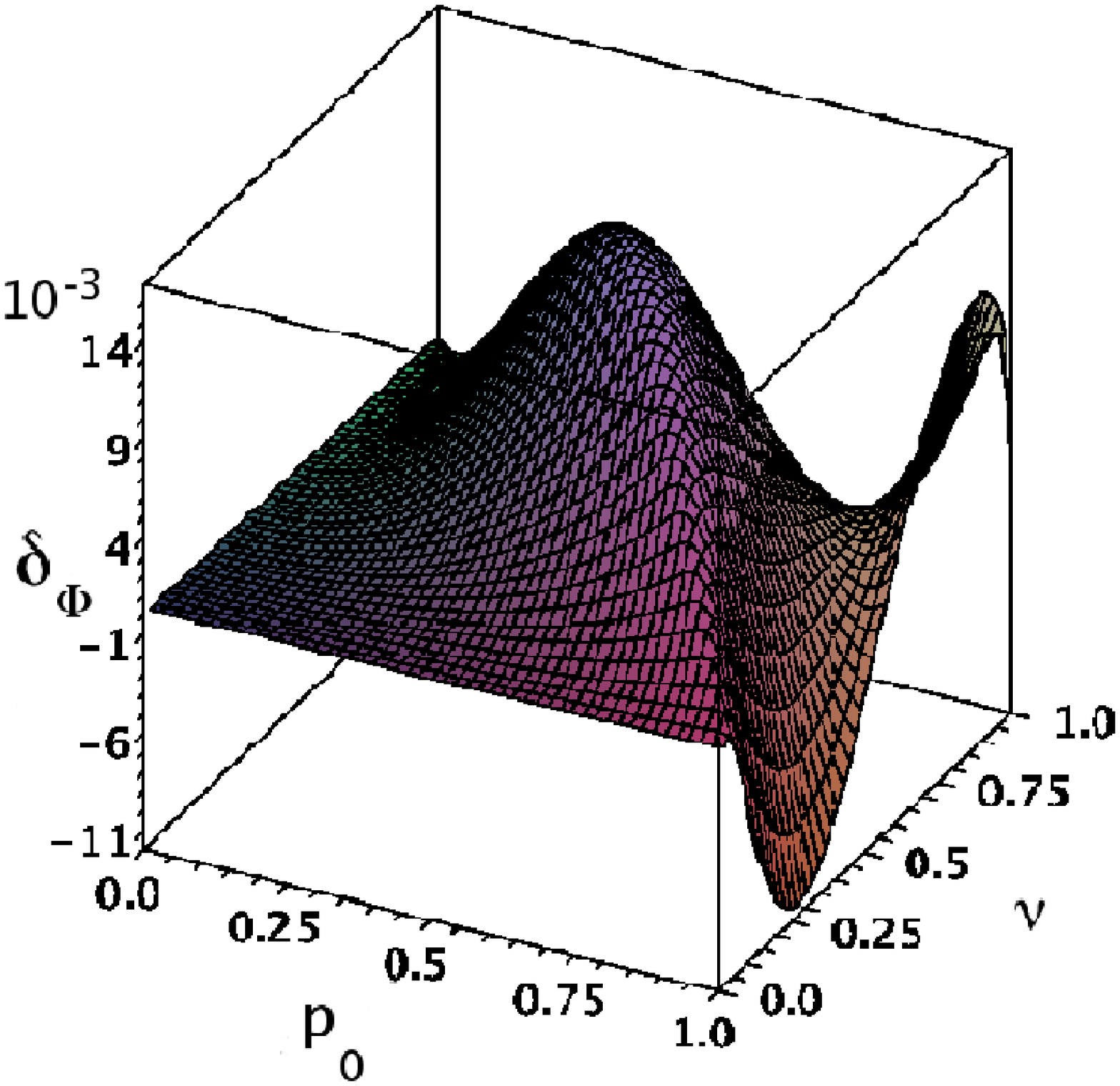}
\caption{This is a plot for $n=2$ of the surface  $\delta_{\Phi}$ for
rays in domain $I$ when rotated to haave the orientation [-70$^{\circ}$, 64$^{\circ}$].}
\vspace{0.4cm}
\label{fig5}
\end{center}
\end{figure}

This observation implies that one can approximate $\Phi$ by the
following expression
\be\n{delta}
\Phi\approx \Phi^a=\hat{\Phi}+\Delta \hat{\Phi}\, .
\ee
Here $\hat{\Phi}$ is given by \eq{a}-\eq{c}, and $\Delta \hat{\Phi}$ is given
by \eq{DP1}-\eq{DP2}.
The relative error of this approximation is
\be
\delta_{\Phi}={\Phi-\Phi^a\over \Phi}
={\Delta\Phi -\Delta \hat{\Phi}\over \Phi}\, .
\ee
Such a relative error calculated in the domain $I$ belongs to some
error interval. These error intervals  for the spacetimes with
different number of dimensions are given in Table~\ref{tt1}. 
The figure~5  illustrates how the relative error $\delta_{\Phi}$ is
distributed in the space of parameters $(p_0,\nu)$. This particular
plot  of the surface $\delta_{\Phi}$ is again shown for $n=2$.
The plots for other values of $n$ are similar. It is evident from the figure
that our approximation for the bending angle works very well
everywhere in the domain $I$ including the region near
$(p_{0}=1, \nu =1)$ which is precisely where the integral expression for the
bending angle diverges.

\subsection{Type $II$ rays}

\subsubsection{Leading part}

For the light ray emitted from the domain $II$ the main contribution to the bending angle is
from an interior point of the integration domain. For this reason the
above procedure for obtaining the analytical approximation for the
bending angle must be slightly modified. We rewrite the
expression \eq{bend} for the bending angle in the form
\be\n{P2}
\Phi=\mu\sqrt{n+2\over n}\int_0^{p_0}{dp\over Z(p)}\, ,
\ee
where, as earlier $p=q/q_*$, $\mu=l/l_*$ and
\be
Z^2(p)=1-{n+2\over n}\mu^2 p^2(1-{2\over n+2}p^n)\, .
\ee
Since $p_0>1$ and the function $Z(p)$ has its minimum at $p=1$, it is
convenient to split the integration domain in \eq{P2} into intervals
$(0,1)$ and $(1,p_0)$. To approximate the first integral
\be
\Phi_*=\mu\sqrt{n+2\over n}\int_0^{1}{dp\over Z(p)}
\ee
one can use the expression obtained in the previous section
\be
\Phi_*\approx\hat{\Phi}_*+\Delta\hat{\Phi}_*\, ,
\ee
where $\hat{\Phi}_*$ is given by \eq{PPP}, while $\Delta\hat{\Phi}_*$
is  $\Delta\hat{\Phi}$, given by \eq{DP1}, calculated for $p_0=1$. 
Thus, for our purposes, it is sufficient to discuss the approximation
of the following quantity
\be\n{PS}
\Phi_{II}=\mu\sqrt{n+2\over n}\int_1^{p_0}{dp\over Z(p)}\, .
\ee

Expanding the function $Z^2(p)$ in the powers of  $(p-1)$ near its
minimum, $p-1=0$, and keeping the terms up to the second order one
obtains
\be\n{ZZZ}
\hat{Z}^2= (1-\mu^2)+(n+2)\mu^2 (p-1)^2\, .
\ee
As earlier, we substitute $\hat{Z}(p)$ instead of $Z(p)$ in
\eq{PS}
\be
\hat{\Phi}_{II}=\mu\sqrt{n+2\over n}\int_1^{p_0}{dp\over \hat{Z}(p)}\, .
\ee

This integral can be taken explicitly and defining { \bf
$z=\mu\sqrt{n+2}(p_0-1)$ } one gets
\ba
\hat{\Phi}_{II}&=&
{1\over \sqrt{n}}\ln{ z
+\sqrt{z^2+1-\mu^2}\over \sqrt{1-\mu^2}}\, .
\ea

We now turn our attention to the  relative error in the approximation
of $\Phi_{II}$ by $\hat{\Phi}_{II}$. The relative error is
\be
\delta^{\Phi}_{II}={\Phi_{II}-\hat{\Phi}_{II}\over \Phi_{II}}
={\Delta{\Phi}_{II}\over \Phi_{II}}\, .
\ee
The relative errors in the dimensions considered are given in Table
II and the figure~6 is a plot of $\delta^{\Phi}_{II}$ in five
dimensions, the plots for other dimensions are similar.   One can
improve the accuracy of the approximation for $\Phi_{II}$ as it was
done for rays emitted in the domain $I$. However, for most of our purposes it
is sufficient to approximate $\Phi_{II}$ by $\hat{\Phi}_{II}$. This
happens because the relative contribution to the bending angle by the
inner part of the ray lying within the domain $II$, $(1>q>q_*)$, is
smaller than the contribution from the part of this ray in the domain
$I$. For this reason the contribution to the `inaccuracy' of the inner
part to the total `inaccuracy' is suppressed. This can be seen
from the Table II which contains both relative errors,
$\delta_{\Phi}$ and $\delta^{\Phi}_{II}$.  Thus  for the
approximation of the inner contribution $\Phi_{II}$ to  the bending
angle it is sufficient to use only the leading term
$\hat{\Phi}_{II}$. 

Using this information we approximate the bending angle for light
propagating to infinity from domain $II$ by
\be
\Phi^{a}=\hat{\Phi}_{*}+\Delta\hat{\Phi}_{*}+\hat{\Phi}_{II}\, .
\ee
The use of our approximation in the previous section and
$\hat{\Phi}_{II}$ accounts for the reasonably good  relative error of
our approximation of the bending angle for light coming from domain
$II$. The relative error of our approximation is
\be
\delta_{\Phi}={\Phi-\Phi^{a}\over \Phi}\, .
\ee
The Table $II$ contains the relative errors appropriate to each value
of $n$ considered. The figure~7 is a plot of $\delta_{\Phi}$ for
$n=2$. We can see that our approximation remains good in the region
where the bending angle diverges.

\begin{figure}[tp]
\begin{center}
\includegraphics[height=2.5in]{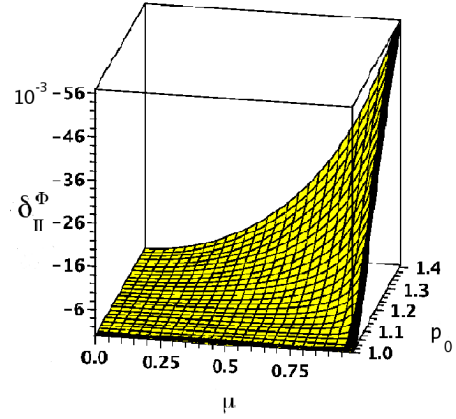}
\caption{This is a plot for $n=2$ of the surface $\delta^{\Phi}_{II}$ for
rays in domain $II$ for $p_0$ satisfying $1\le p_0 \le 1.41$, $0\le \mu \le0.99$ when it
is rotated to have an orientation [-9$^{\circ}$, -107$^{\circ}$]. We don't show the region for $0.99 \le \mu \le 1$ because the plot descends steeply to zero in this region and it is difficult to show the nature of the plot of this region on this scale.}
\label{fig6}
\end{center}
\end{figure}

\begin{table}[htb]
\begin{center}
\begin{tabular}{|l|c|c|c|c|c|}
\hline
Dimension & $p_0$ & Percentage error $\delta^{\Phi}_{II}$ & Percentage error
$\delta_{\Phi}$
\\[1.0ex]
\hline
D=4,n=1 & $p_0=1.49$ & [-4.38\%, 0.00\%] & [-1.30\%, 0.83\%] \\[1.0ex]
\hline
D=5,n=2 & $p_0=1.41$ & [-5.69\%, 0.00\%]  & [-1.80\% , 1.15\%]  \\[1.0ex]
\hline
D=6,n=3 & $p_0=1.35$ & [-6.64\%, 0.00\%]  & [-2.11\% , 1.72\%]  \\[1.0ex]
\hline
D=7,n=4 & $p_0=1.31$ & [-7.50\%, 0.00\%]  & [-2.25\% , 2.56\%]  \\[1.0ex]
\hline
\end{tabular}
\end{center}
\caption{Above are the relative errors associated with the quantities
$\delta^{\Phi}_{II}$ and $\delta_{\Phi}$ for $\Phi_{II}$ and $\Phi$
in the domain $II$ for $p_0$ satisfying $1\le p_0 \le q^{-1}_{*}$ for
the spacetimes of dimension $D=n+3$. Also included are the maximal values of
$p_0$ considered for the endpoint of integration.} \n{tt2}
\end{table}

\begin{figure}[tp]
\begin{center}
\includegraphics[height=2.5in]{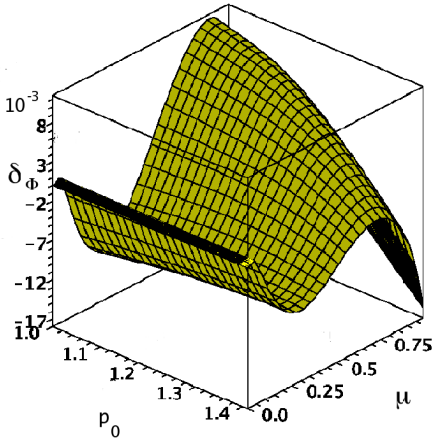}
\caption{This is a plot for $n=2$ of the surface $\delta_{\Phi_{II}}$ for
rays in domain $II$ for $p_0$ satisfying $1\le p_0 \le 1.41$,$0\le \mu \le0.99$ when it
is rotated to have an orientation [-48$^{\circ}$, 62$^{\circ}$]. We don't show the region for $0.99 \le \mu \le 1$ because the plot descends steeply to zero in this region and it is difficult to show the nature of the plot of this region on this scale.}
\vspace{0.4cm}
\label{fig7}
\end{center}
\end{figure}

\section{Approximating the time delay}

\subsection{Type $I$ rays}

\subsubsection{Leading part}

The time delay $t^{(o)}-t^{(e)}$, \eq{to}, and the corresponding delay
in the retarded time $u$, \eq{uo}, is defined by the function $T$,
\eq{TT}, or its dimensionless form $\tau$
\ba\n{T}
T=r_{g}q^{-1}_{0} \tau\hh
\tau=\int^{1}_{0}\frac{dy}{(1-y)^2\, f}
\left( {\sqrt{f_0}\over Q}-1\right)\, .\n{tau}
\ea

We now construct
an analytic approximation for $\tau$ in the same way as it was done
for $\Phi$ in the previous section. It is easy to see that in
spite of the factor $(1-y)^2$ in the denominator of the integrand in
\eq{tau}, the integral is finite at $y=1$. The only singular point of
$\tau$ in the domain $I$ is its logarithmic divergence at $y=0$
for the limiting values of the parameters $p_0=\nu=1$. To extract
this divergence we subtract from $\tau$ a similar integral, where
instead of $Q$ we use $\hat{Q}$ given by \eq{QQ}. In order to be able
to obtain the value of the integral in an explicit form in terms of
elementary functions, as earlier, we make another modification of the
integrand. Namely, instead of $H=(1-y)^2f$ we use its quadratic
expansion near $y=0$ of the form
\ba\n{HHH}
\hat{H}&=&f_{0} +2\beta y+\gamma y^2\,
\ea
\ba
f_{0} &=&1-{2P\over n+2}\, ,\quad \beta =-(1-P)\, ,\\
\gamma &=&1-(n+1)P\hhh P=p_0^n\,.
\ea

 It is easy to see that $f_{0}\ge 0$ and that the discriminant of the
quadratic in $y$, \eq{HHH}, is of the form  
\ba\n{mu}
f_{0}\gamma-\beta^2=-{nP(n+1-P)\over n+2}={D-f_{0}c\over \nu^4}\, 
\ea
and is non-positive. The parameters $a$, $b$, and $c$, which
enter the expression $\hat{Q}$, \eq{QQ}, are related to the coefficients in
\eq{HHH} as follows
\be\n{L}
a=(1-\nu^2)f_{0} \hhh b= -\nu^2\beta\hhh c = -\nu^2\gamma\, .
\ee
Let us emphasize that the structure of the logarithmic divergence of
the exact integral \eq{tau} at $y=0$ is preserved in our approximation
since $\hat{H}(y=0)=f_{0}=[(1-y)^2f]|_{y=0}$.

We define the quantity $\hat{\tau}$ as follows
\ba\n{P1}
\hat{\tau}=\int^{1}_{0}\frac{dy}{\hat{H}} 
\left( {\sqrt{f_0}\over \hat{Q}}-1\right)\, .
\ea
We note that  $\hat{H}$ has a root in the interval $[0,1]$. Rewriting the integrand in \eq{P1} as follows
\ba
\frac{1}{\hat{H}}
\left( {\sqrt{f_0}\over \hat{Q}}-1\right) = \frac{\nu^2}{\sqrt{f_{0}}\hat{Q}+\hat{Q}^2}\, ,
\ea
one can see that it does not result in a divergence.
Defining the following quantities:
\ba
M &=& f_{0}c-D\nonumber\, ,\\
\cal{A}_{\pm}&=&D+c\sqrt{f_{0}A}\pm\,(c+b)\sqrt{M}\, ,\\
\cal{B}_{\pm}&=&D+c\sqrt{f_{0}a}\pm\,b\sqrt{M}\nonumber\, ,
\ea
where, as earlier, $D= ac-b^2$, the explicit form of $\hat{\tau}$ is
\ba
\hat{\tau}&=&\frac{\nu^2}{2\sqrt{M}}
\ln{\left(\frac{{\cal A}_{+} {\cal B}_{-}}{{\cal A}_{-} {\cal B}_{+}}\right)} .
\ea
In the case $c = 0$, {\bf $(B\neq 0)$},  $\hat{\tau}$ has the following form
\ba
\hat{\tau}&=&\frac{\nu^2}{2b}\ln{\left(\frac{a+2b-f_{0}}{a-f_{0}}\right)}
\nonumber\,\\
&+&\frac{\nu^2}{b}
\tanh^{-1}\left(\frac{\sqrt{f_{0}}(\sqrt{a+2b}-\sqrt{a})}
{f_{0}-\sqrt{a}\sqrt{a+2b}}\right)\, .
\ea

For the boundary of the domain $I$ where $q=q_*$ ($p_0=1$) the
expression for $\hat{\tau}$ simplifies and takes the form
\ba\n{www}
\hat{\tau}_* & = & {\sqrt{n+2}\over 2n} \ln{(Z_+/Z_-)}\, ,\\
Z_{\pm} & = & Bn\sqrt{n+2}\pm (n-nB+n\sqrt{(n+1)B+1})\, .\nonumber
\ea
For the near critical rays ($\mu-1\approx 0$)
\be
\hat{\tau}_*\sim {\sqrt{n+2}\over 2n}\left[ -\ln (\mu-1)+\ln \left(
{{n+2}\over 2\sqrt{n+2}+n+3}\right)\right]+\ldots\, .
\ee
This time is logarithmically divergent at $\mu\to 1$.

As in the case for $\Phi$ we can express the quantity $\tau$ as 
\ba
\tau=\hat{\tau}+\Delta\tau\, ,
\ea
the quantity $\Delta\tau$ being finite for rays moving along critical
trajectories.

\subsubsection{Approximating $\Delta \tau$}

\begin{figure}[tp]
\begin{center}
\includegraphics[height=2.8in]{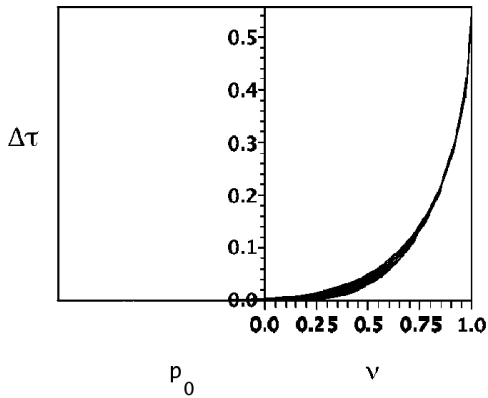}
\caption{This is a plot for $n=2$ of the surface $\Delta{\tau}$ when
it is rotated to have an orientation [-45$^{\circ}$, 90$^{\circ}$].
We note that we use the orientation [-43$^{\circ}$, 90$^{\circ}$] in
our approximation because it better approximates a curve.}
\vspace{0.4cm}
\label{fig8}
\end{center}
\end{figure}

As earlier, by studying of the plots for $\Delta{\tau}$ as a function of 2
variables $p_0$ and $\nu$ one can observe that they can be
approximated by a function of one variable, which is a linear
combination of $p_0$ and $\nu$. For example, figure~\ref{fig8}  shows
the 3D plot of $\Delta\tau$ for $n=2$ ($D=5$) with the orientation
option chosen to be $[-43^{\circ}, 90^{\circ}]$. The 2D surface in
this projection is again a slightly broadened curve. This allows one 
to approximate $\Delta\tau$  with a very good accuracy by a
function 
\be
\Delta{\hat{\tau}}=-k_{\tau}p_{0}\nu(\sqrt{1-(x_{\tau}-1)^2}-1)\Theta(x_{\tau}-1)\,
, 
\ee 
\be 
x_{\tau}=\sqrt{2}[\cos(\frac{\pi}{4}+\psi_{\tau})
p_{0}+\sin(\frac{\pi}{4}+\psi_{\tau})\nu] \, . 
\ee
The corresponding values of $k_{\tau}$ and $\psi_{\tau}$
are given in Table~\ref{tt3}. Thus we can approximate $\tau$ as
follows
\be\n{tap}
\tau\approx \tau^a=\hat{\tau}+\Delta{\hat{\tau}}\, .
\ee

\begin{table}[htb]
\begin{center}
\begin{tabular}{|l|c|c|c|c|c|}
\hline
Dimension & $k_{\tau}$  & $\psi_{\tau}$ & Percentage error
$\delta_{\tau}$
\\[1.0ex]
\hline
D=4,n=1  & $0.66$ & $\pi/45$ &[-1.68\% , 2.71\%] \\[1.0ex]
\hline
D=5,n=2 & $0.55$  & $\pi/90$ &[-2.19\% , 2.60\%]  \\[1.0ex]
\hline
D=6,n=3 & $0.52$  & $0$ &[-3.28\% , 2.31\%]  \\[1.0ex]
\hline
D=7,n=4  & $0.52$ & $\pi/60$ &[-1.73\% , 2.66\%]  \\[1.0ex]
\hline
\end{tabular}
\end{center}
\caption{Approximation parameters $k_{\tau}$ and
$\psi_{\tau}$, and range of relative errors $\delta_{\tau}$ for $\tau$
in the domain $II$ for the spacetime of dimension $D=n+3$.}
\n{tt3}
\end{table}

\begin{figure}[tp]
\begin{center}
\includegraphics[height=2.5in]{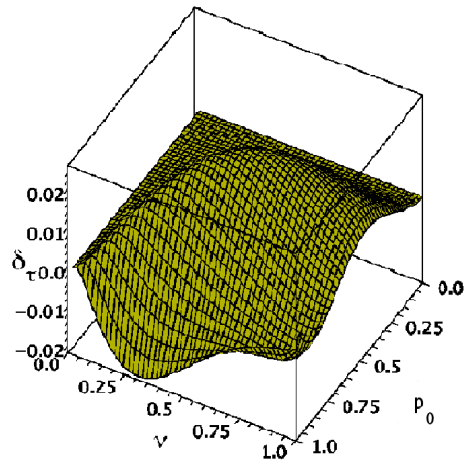}
\caption{This is a plot for $n=2$ of
the surface $\delta_{\tau}$ when it is rotated to have an orientation
[29$^{\circ}$, 46$^{\circ}$]}
\vspace{0.4cm}
\label{fig9}
\end{center}
\end{figure}

The figure~9 is a plot of the surface $\delta_{\tau}$ for the domain
$I$, $0\le \lbrace p_{0}, \nu \rbrace\le 1$. Here we have shown
$\delta_{\tau}$ for $n=2$. The plots for other values of $n$ are
similar.  The relative error of this approximation is
\be
\delta_{\tau}={\tau-\tau^a\over \tau}
={\Delta\tau -\Delta{\hat{\tau}}\over \tau}\, .
\ee
The relative errors $\delta_{\tau}$ for a spacetime with different
number of dimensions are given in Table~\ref{tt3}. It is evident from
the figure that our approximation for the time delay works very well
in the region near $(p_{0}=1, \nu =1)$ which is precisely where the
integral expression for the retarded time diverges.

\subsection{Type $II$ rays}

\subsubsection{Leading part}

In this section we investigate the retarded time for light rays that
proceed to infinity from the domain $II$ where $p_0$ satisfies $1 \le
p_0 < ((n+2)/2)^{1/n}$. 
As in the case of type $I$ rays we express the retarded time for a
light ray emitted in domain $II$ in terms of dimensionless quantities. In
this instance we use the quantities $p_0$ and $\mu$ as defined
earlier. After reformulating the expression for the retarded time, \eq{TT}, in
terms of these variables we have
\ba\n{jj}
 &&T=r_g q_*^{-1}\tau\hhh
\tau=\mu^2 \left(1+{2\over n}\right) J(0,p_0)\, ,\\
&&J(p_1,p_0)=\int^{p_0}_{p_1}
\frac{{dp}}{Z(p)(1+Z(p))}\, .
\ea
Since we are in a region where $p_0 >1$ and $Z(p)$  has a minimum at
$p=1$ we again split the integration domain in $\eq{jj}$ into two
intervals, $(0,1)$ and $(1,p_0)$. In the first interval we can write
the contribution to the retarded time as 
\ba
\tau_{*}=\mu^2 \left(1+{2\over n}\right) J(0,1)\, .
\ea
This integral can be approximated by
\ba
\tau_{*}\approx\hat{\tau}_{*}+\Delta\hat{\tau}_{*}\, ,
\ea  
where $\hat{\tau}_{*}$ is given by $\eq{www}$ and
$\Delta\hat{\tau}_{*}$ is $\Delta\hat{\tau}$ evaluated at $p_0=1$.
Therefore in approximating the retarded time in domain $II$ it is
sufficient for our purposes to consider the following integral
\be
\tau_{II}=\mu^2 \left(1+{2\over n}\right) J(1,p_0)\, .
\ee
Identically to what we did earlier we substitute $\hat{Z}(p)$ given by
\eq{ZZZ}  for $Z(p)$ to obtain
\ba
&&\hat{\tau}_{II}=\mu^2 \left(1+{2\over n}\right) \hat{J}(1,p_0)\, ,\\
&&\hat{J}(1,p_0)=\int^{p_0}_{p_1}
\frac{{dp}}{\hat{Z}(p)(1+\hat{Z}(p))}\, .
\ea
This integral is calculated explicitly using { \it Maple} and is given by
\ba
\hat{\tau}_{II}=\frac{\sqrt{n+2}}{2n}
\ln{\frac{\sqrt{1-\mu^2+z^2}+1-\mu^2+\mu z}
{\sqrt{1-\mu^2+z^2}+1-\mu^2-\mu z}}\, .
\ea

We now focus our attention on the relative error $\delta^{\tau}_{II}$
\ba
\delta^{\tau}_{II}={\tau_{II}-\hat{\tau}_{II}\over \tau_{II}}
={\Delta\hat{\tau}_{II}\over \tau_{II}}\, 
\ea
of the approximation of $\tau_{II}$ by $\hat{\tau}_{II}$.   In Table
IV we present the relative errors of our first order approximation of
$\tau_{II}$ for dimensions $4, 5, 6, 7$. The figure~10 is a plot of
$\delta^{\tau}_{II}$ in five dimensions, the plots in other
dimensions are similar.  As earlier, examining the results presented
in Table IV we can conclude that  $\hat{\tau}_{II}$ provides a good
first order approximation of the function $\tau_{II}$  in the domain
$II$. Therefore in approximating the retarded time for light going to
infinity from domain II we use
\ba
\tau^{a}=\hat{\tau}_{*}+\Delta\hat{\tau}_{*}+\hat{\tau}_{II}\, ,
\ea 
and the corresponding relative error of our approximation for the
retarded time delay is 
\ba
\delta_{\tau}={\tau-\tau^{a}\over\tau}\, .
\ea
This relative error for
each dimension considered is shown in Table IV. The figure~11 is a
plot of $\delta_{\tau}$ for $n=2$. We can see that our approximation
is very good in the region $\mu\sim1$, which is where the retarded
time diverges.

\begin{figure}[tp]
\begin{center}
\includegraphics[height=2.5in]{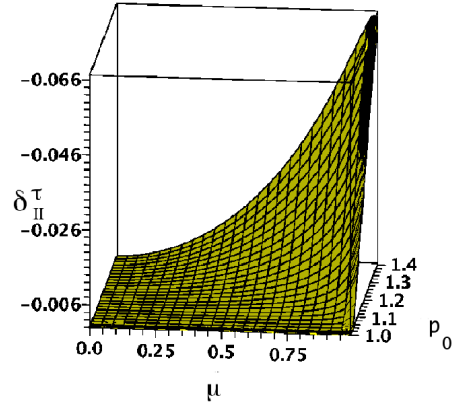}
\caption{This is a plot for $n=2$ of the surface $\delta^{\tau}_{II}$ in
domain $II$ for $p_0$ satisfying $1\le p_0 \le 1.41$, $0\le \mu \le0.99$ when it is
rotated to have an orientation [-6$^{\circ}$, -106$^{\circ}$]. We don't show the region for $0.99 \le \mu \le 1$ because the plot descends steeply to zero in this region and it is difficult to show the nature of the plot of this region on this scale.}
\vspace{0.4cm}
\label{fig10}
\end{center}
\end{figure}

\begin{table}[htb]
\begin{center}
\begin{tabular}{|l|c|c|c|c|c|}
\hline
Dimension & $p_0$  & Percentage error $\delta^{\tau}_{II}$ &
Percentage error $\delta_{\tau}$
\\[1.0ex]
\hline
D=4,n=1& $p_0=1.49$  & [-5.18\%, 0.00\%] &[-2.00\%, 0.72\%] \\[1.0ex]
\hline
D=5,n=2 & $p_0=1.41$ & [-6.73\%, 0.00\%] &  [-2.52\%, 0.16\%]  \\[1.0ex]
\hline
D=6,n=3 & $p_0=1.35$ & [-7.83\%, 0.00\%] &  [-3.24\%, 0.67\%]  \\[1.0ex]
\hline
D=7,n=4 & $p_0=1.31$ & [-8.83\%, 0.00\%] & [-2.59\%, 2.67\%]  \\[1.0ex]
\hline
\end{tabular}
\end{center}
\caption{This table shows the relative errors associated with the
quantities $\delta^{\tau}_{II}$ and $\delta_{\tau}$ for  our
approximation of $\tau_{II}$ and $\tau$  in the domain $II$ for $p_0$
satisfying $1\le p_0 \le q^{-1}_{*}$ for the spacetimes of dimension
$D=n+3$. Also included are the maximal values of $p_0$ considered for the endpoint
of integration.} \n{tt4} 
\end{table}

\begin{figure}[tp]
\begin{center}
\includegraphics[height=2.5in]{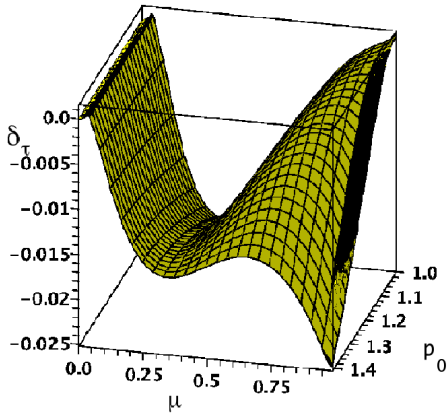}
\caption{This is a plot for $n=2$ of the surface $\delta_{\tau}$ for
rays in domain $II$ for $p_0$ satisfying $1\le p_0 \le 1.41$,$0\le \mu \le0.99$ when it
is rotated to have an orientation [14$^{\circ}$, 67$^{\circ}$]. We don't show the region for $0.99 \le \mu \le 1$ because the plot descends steeply to zero in this region and it is difficult to show the nature of the plot of this region on this scale.}
\vspace{0.4cm}
\label{fig11}
\end{center}
\end{figure}

\section{Summary and Discussion}

In the present paper we have investigated ray-tracing problem for
photons propagating in four and higher dimensional spherically
symmetric black hole backgrounds. We focused on the  bending angle
and the retarded time for a trajectory of the null ray with the impact
parameter $\lambda$ which starts at some finite radius $r_0$ and
propagates to the infinity. We obtained an analytical expression in
terms of the elementary functions which uniformly approximate the
bending angle $\Phi_{\lambda}(r_0)$ and the retarded time function
$T_{\lambda}(r_0)$ with high accuracy. Knowledge of these two
functions, $\Phi_{\lambda}(r_0)$ and  $T_{\lambda}(r_0)$, allows one
to calculate other quantities which are of interest in possible applications. 

For example, let us consider a scattering problem, when a photon with
a given impact parameter $\lambda$ comes from infinity and after
passing near the black hole goes to infinity again. The total bending
angle for this process is $2\Phi_{\lambda}(r^*)$, where $r^*$ is the
radius of the photon's turning point. This quantity can be
approximated by $2\hat{\Phi}_*$, where $\hat{\Phi}_*$ is given by
\eq{PPP}. Similarly, the expression \eq{www} can be used for an
approximation of the time delay quantities for the scattering problem.

A knowledge of the bending angle function $\Phi_{\lambda}(r)$ is
sufficient for a  reconstruction a complete ray
trajectory. Really, let us fix the impact parameter $\lambda$ and consider
a ray with this impact parameter emitted at 
the radius $r_0$ in the D-dimensional Schwarzschild-Tangherlini
space. Since the ray trajectories are planar, it is sufficient to
consider rays lying in the equatorial plane passing through the point
of emission. We choose $\phi_0=0$ for this point.
The ray trajectory is defined by the equation
\be
\phi=F(r)=\Phi_{\lambda}(r)-\Phi_{\lambda}(r_0)\, .
\ee
To obtain an approximated form of the
trajectory equation it is sufficient to substitute $\Phi^a$ instead
of the exact values $\Phi$. 

Similarly one can use the approximating expression for solving the
problem of the reconstruction of a null geodesic connecting two
points $(r_0,\phi_0)$ and $(r_1,\phi_1)$. In the 4 dimensional case
this problem can be reduced to one non-linear equation which contains
as the arguments the Jacobi elliptic functions \cite{CaKo}. By using
the developed approximation one can reduce this problem in any number
of dimensions to solving non-linear equations which contain only
elementary functions.

We hope that the obtained results will be useful in studying these and other
problems connected with light propagation in the black hole
vicinity which are of interest in the application to astrophysics.
Since the developed approximation is valid in a spacetime with
arbitrary number of dimensions, it might be useful also for study
black holes in the models with large extra dimensions.

\medskip

\noindent 

\section*{Acknowledgments} 

\noindent 
One of the authors, V. F., would like to thank the Natural Sciences
and Engineering Research Council of Canada (NSERC) and the Killam
Trust for financial support. The other author, P.C., would like to
thank the Department of Physics at the University of Alberta for continued financial assistance.

\end{document}